\documentstyle[11pt]{article}\textheight 230mm\textwidth 150mm
            \newcommand{\be}{\begin{eqnarray}}
            \newcommand{\ee}{\end{eqnarray}}
            \newcommand{\eel}[1]{\label{#1}\end{eqnarray}}
\newcommand{\e}[1]{\label{e:#1}\end{eqnarray}}
     \newcommand{\eg}{{\em e.g.\ }}
            \newcommand{\ie}{{\em i.e.\ }}

            \newcommand{\la}{{\lambda}}
            
            \newcommand{\del}{{\delta}}

\newcommand{\cS}{{\cal{S}}}
\newcommand{\cA}{{\cal{A}}}
\newcommand{\cM}{{\cal{M}}}
\newcommand{\cU}{{\cal{U}}}

 \newcommand{\lea}{{\leftarrow}}
            \newcommand{\lra}{{\leftrightarrow}}
            \newcommand{\Lra}{{\Leftrightarrow}}

            \newcommand{\beq}{\begin{quote}}
            \newcommand{\eq}{\end{quote}}

            \newcommand{\al}{\alpha}
            \newcommand{\ben}{\begin{enumerate}}
            \newcommand{\een}{\end{enumerate}}
            \newcommand{\bit}{\begin{itemize}}
            \newcommand{\ei}{\end{itemize}}
    	\newcommand{\nn}{\nonumber}
            \newcommand{\r}[1]{(\ref{e:#1})}
            \newcommand{\edfl}[1]{\label{#1}\end{df}}
\newcommand{\vb}{{\cal h}}
\newcommand{\hb}{{\cal i}}

\newcommand{\ve}{{\varepsilon}}

\newcommand{\bett}{{\bf 1}}
	
\def\d{\partial}

  \def\half{{1 \over 2}}

\def\JMP{{\sl J.\ Math.\ Phys.}}
\begin{document}
\begin{titlepage}
\noindent
G\"{o}teborg ITP 98-04\\

\vspace*{5 mm}
\vspace*{35mm}
\begin{center}{\LARGE\bf Quantum antibrackets}
\end{center} \vspace*{3 mm} \begin{center} \vspace*{3 mm}

\begin{center}Igor Batalin\footnote{On leave of absence from
P.N.Lebedev Physical Institute, 117924  Moscow, Russia\\E-mail:
batalin@td.lpi.ac.ru.} and Robert
Marnelius\footnote{E-mail: tferm@fy.chalmers.se.}\\
\vspace*{7 mm} {\sl Institute
of Theoretical Physics\\ Chalmers University of
Technology\\ G\"{o}teborg
University\\ S-412 96  G\"{o}teborg, Sweden}\end{center}
\vspace*{25 mm}
\begin{abstract}
A binary expression in terms of operators is given which satisfies all the
quantum counterparts of the algebraic properties of the classical antibracket.
This quantum antibracket has therefore the same relation to the classical
antibracket as commutators to Poisson brackets. It is explained how this quantum
antibracket is related to the classical antibracket and the $\Delta$-operator in
the BV-quantization. Higher quantum antibrackets are introduced in terms of
generating operators, which automatically yield all their subsequent Jacobi
identities as well as the consistent Leibniz' rules.
\end{abstract}\end{center}\end{titlepage}

\setcounter{page}{1}

In classical dynamics there are two basic binary operations, one is the Poisson
bracket and the other is the antibracket \cite{anti,BV}. The latter has mainly
been used in the BV-quantization of gauge theories \cite{BV}. In canonical
quantization classical dynamical functions are mapped on operators and Poisson
brackets are mapped on commutators. A long standing problem has been to
understand what the quantum counterpart of the classical antibracket is. This
problem as well as the related problem of coexistence of Poisson brackets and
antibrackets have \eg been considered in
\cite{Panti,KN,quanti,AMS}. Here we give a
natural solution to this problem, a solution which will both deepen our
understanding of the BV-quantization as well as provide for further
generalizations.

Let us first remind about the mapping
of functions on a given symplectic
manifold to operators. This mapping is such that all defining
properties of the Poisson bracket  is  satisfied by the (super)commutator. The
defining properties of the Poisson bracket, $\{f, g\}$, for functions $f,
g$ on a
manifold $\cS$ are (The Grassmann parities are denoted by
$\ve_f\equiv\ve(f)=0,1$ mod $2$.)\\
\\ i) Grassmann parity
\be
&&\ve(\{f, g\})=\ve_f+\ve_g.
\e{1}
ii) Symmetry
\be
&&\{f, g\}=-\{g, f\}(-1)^{\ve_f\ve_g}.
\e{2}
iii) Linearity
\be
&&\{f+g, h\}=\{f, h\}+\{g, h\}, \quad (\ve_f=\ve_g).
\e{3}
iv) Jacobi identities
\be
&&\{f,\{g, h\}\}(-1)^{\ve_f\ve_h}+cycle(f,g,h)\equiv0.
\e{4}
v) Leibniz' rule
\be
&&\{fg, h\}=f\{g, h\}+\{f, h\}g(-1)^{\ve_g\ve_h}.
\e{5}
vi) For any odd/even parameter $\la$ we have
\be
&&\{f, \la\}=0\quad {\rm any}\;f\in\cS.
\e{6}
The Poisson bracket can only be nondegenerate if the dimension of the even
subspace of $\cS$ is even. ($\cS$ is then a symplectic manifold.)

The quantization is a mapping of all functions $f\in\cS$ to noncommutative
operators $f$ such that the Grassmann parities are preserved. The
(super)commutator of such operators are defined by
($f$ and $g$ are now operators)
\be
&&[f, g]\equiv fg-gf(-1)^{\ve_f\ve_g}.
\e{7}
One may easily check that the commutator satisfies all the properties
i) to vi) with the Poisson bracket replaced by the commutator and the functions
replaced by their corresponding operators. Note that the commutator for classical
functions vanish.

We construct now the  corresponding mapping for
antibrackets. Consider therefore a manifold, $\cA$, on which we have the
binary operation called antibracket.
The defining properties of the antibracket $(f,g)$
for functions $f, g \in\cA$ are\\
\\ 1) Grassmann parity
\be
&&\ve((f, g))=\ve_f+\ve_g+1.
\e{8}
2) Symmetry
\be
&&(f, g)=-(g, f)(-1)^{(\ve_f+1)(\ve_g+1)}.
\e{9}
3) Linearity
\be
&&(f+g, h)=(f, h)+(g, h), \quad (\ve_f=\ve_g).
\e{10}
4) Jacobi identities
\be
&&(f,(g, h))(-1)^{(\ve_f+1)(\ve_h+1)}+cycle(f,g,h)\equiv0.
\e{11}
5) Leibniz' rule
\be
&&(fg, h)=f(g, h)+(f, h)g(-1)^{\ve_g(\ve_h+1)}.
\e{12}
6) For any odd/even parameter $\la$ we have
\be
&&(f, \la)=0\quad {\rm any}\;f\in\cA.
\e{13}
The antibracket can only be nondegenerate if $\cA$ is a supermanifold with
the dimension $(n,n)$. ($\cA$ is then an antisymplectic manifold.)

Consider now a mapping of all functions on $\cA$ to  operators which
preserves the Grassmann parities. (If $\cA$ is of dimension $(2n,2n)$  it may
 be both a symplectic as well as an antisymplectic manifold which could allow
for a canonical quantization.) We look then for a quantum antibracket,
$(\;,\;)_Q$,
satisfying the counterparts of the properties 1)-6).
There are not many candidates for
such a quantum bracket. Condition 1) requires the presence of an odd
operator $Q$
which should be fundamental. It could in principle be a  rather abstract
operator
with no classical counterpart. However, in the
following we treat $Q$ as an ordinary
odd operator. Conditions 2) and
3) require an expression which is linear in the operators entering the quantum
antibracket. Therefore
$(f, g)_Q$ must consists of terms which are simple products of $f$, $g$ and $Q$.
Let us start with the simplest ansatz of this type satisfying 1)-3). It is
\be
&&(f, g)_Q\equiv fQg-gQf(-1)^{(\ve_f+1)(\ve_g+1)}.
\e{14}
A remarkable property of this ansatz  is that it not only satisfies
conditions 1)-3) but also the Jacobi identities 4) without any restrictions
on the
operators. However, in distinction to the corresponding case for the commutator
\r{7}, the properties 5) and 6) are not automatically satisfied by
\r{14}. In fact, they require all operators $f$ to satisfy the
condition\footnote{The ansatz \r{14} satisfies the Leibniz' rule 5) for any
operators provided the operator multiplication is defined by $f\cdot
g\equiv fQg$.
We have then $(f, gQh)_Q=(f,g)_Q\,Qh+gQ(f, h)_Q\,(-1)^{(\ve_f+1)(\ve_g+1)}$.}
\be
&&[Q, f]_+=0,
\e{17}
where
\be
&&[f, g]_+\equiv fg+gf(-1)^{\ve_f\ve_g}.
\e{18}
The condition \r{17}
 restricts the class of
 operators considerably. General solutions seem to be
\be
&&f=[Q, \phi]_+, \quad
[Q^2, \phi]=0.
\e{22}
The ansatz \r{14} is therefore unsatisfactory.

We have to look for a better proposal for the quantum antibracket than
\r{14}. We
consider then the most general ansatz consisting of terms which are products of
$f$, $g$ and $Q$ such that conditions 1)-3) are satisfied. Then we require this
ansatz to satisfy condition 6) and to be such that its classical limit is zero
without any conditions on $f$ or $g$. (The classical limit of \r{14} is
only zero
if one of the operators $f$ or $g$ is of the allowed form \r{22}.) The
solution is
unique up to a factor and is
\be
&&(f, g)_Q=\half \left([f, [Q, g]]-[g, [Q, f]](-1)^{(\ve_f+1)(\ve_g+1)}\right).
\e{24}
After appropriate restrictions this is the correct proposal for the
quantum antibracket in our opinion. Explicitly we have
\be
&&(f, g)_Q=fQg-gQf(-1)^{(\ve_f+1)(\ve_g+1)}-[Q,
\half[f,g]_+]_+(-1)^{\ve_f}.
\e{241}
The first two terms is just the ansatz \r{14} and the last two terms is a
double symmetrization according to \r{18}.
One may note that both \r{14} and
\r{24} satisfy
$(Q, Q)_Q\equiv0$.   For operators $f, g$,
commuting with $Q^2$ \r{24} satisfies $(f,
Q)_Q=0$ and
\be
&&[Q, (f, g)_Q]=([Q, f], g)_Q+(f, [Q,
g])_Q(-1)^{\ve_f+1}=[[Q,f], [Q, g]].
\e{242}
Now the expression \r{24} does not automatically satisfy the Jacobi
identities 4)
and the Leibniz' rule 5). Instead of the latter we have
\be
&&(fg, h)_Q-f(g, h)_Q-(f,
h)_Qg(-1)^{\ve_g(\ve_h+1)}=\nn\\
&&=\half\left([f, h][g,
Q](-1)^{\ve_h(\ve_g+1)}+[f,Q][g,h](-1)^{\ve_g}\right).
\e{25}
Therefore condition 5) requires us  to restrict the class of allowed
operators. Eq.\r{25}  suggests two natural possibilities: We could restrict the
class of operators to those which satisfy $[Q, f]=0$. However, this is obviously
very bad since this would again imply that $(f, g)_Q=0$ for allowed
operators. The
second natural alternative is to restrict ourselves to the class of all
commuting
operators. A nontrivial quantum antibracket is then possible if $Q$ does not
belong to this class. (A more general class of operators is considered in
Appendix B.)  For the class of commuting operators the expression
\r{24} simplifies to
\be
&&(f, g)_Q=[f, [Q, g]]=[[f, Q], g], \quad [f, g]=0,
\e{26}
by means of the Jacobi identities for the (super)commutator \r{7}.
The Jacobi identities 4) for the ansatz \r{26} with commuting operators
$f,g,h,$ require the condition
\be
&&[[f, [g, [h,Q]]], Q]-[f, [g, [h, Q^2]]]=0.
\e{271}
This  condition may be viewed as a condition  on $Q$ or as a further restriction
on the class of allowed operators. In the following we choose to view
it as a condition on $Q$, since the class of commuting operators  is a
very natural one to consider. Eq.\r{271} may be split into the following two
slightly stronger conditions (we denote the class of commuting operators $\cM$
from now on)
\be
&&[f, [g, [h, Q]]]=0, \quad \forall f,g,h\in\cM,
\e{251}
and
\be
&&[Q^2, f]=0, \quad \forall f\in\cM\quad \Leftrightarrow\quad Q^2\in\cM.
\e{252}
If $Q$ satisfies conditions \r{251}, \r{252} then \r{26} satisfies all the
required properties of a quantum antibracket for the class of commuting
operators
$\cM$. The condition \r{251} is a very natural condition which is equivalent to
the requirement
\be
&&(f, g)_Q\in\cM, \quad \forall f,g\in\cM,
\e{253}
a condition which means that the quantum antibracket  of commuting operators
is again a commuting operator belonging to $\cM$.  The correspondence to
the classical antibracket is
\be
&&(i\hbar)^{-2}(f, g)_Q\; \lra\; (f, g),
\e{31}
due to the commutator correspondence
\be
&&(i\hbar)^{-1}[f, g]\; \lra\; \{f, g\}.
\e{30}
The expression \r{26} for the quantum antibracket implies  the following
relation between classical antibrackets and Poisson brackets
\be
&&(f, g)=\{\{f, Q\}, g\}=\{ f, \{Q, g\}\}
\e{32}
for all functions $f, g$ satisfying $\{f, g\}=0$. Here
the odd function $Q$ satisfies the classical counterpart of
\r{251} and is such that $\{Q, Q\}$ has zero Poisson brackets with all functions
$f, g$.

The condition
\r{251} when viewed as a condition on Q is rather severe. However, for a general
$Q$ which does not satisfy
\r{251} for arbitrary commuting operators we still have a consistent
scheme in terms of higher quantum antibrackets. This scheme is most conveniently
presented in terms of the generating operator
\be
&&Q(\phi)\equiv
\exp{\{-(i\hbar)^{-1}f_a\phi^a\}}\,Q\,\exp{\{(i\hbar)^{-1}f_a\phi^a\}},
\quad [Q^2, f_a]=0,\;\forall f_a,
\e{321}
where $f_a$ are commuting operators in the set $\cM$ and where $\phi^a$ are
parameters. The Grassmann parities are $\ve(f_a)\equiv\ve_a=\ve(\phi^a)$. Since
$Q^2$ commutes with $f_a$ we have
\be
&&[Q(\phi), Q(\phi)]=[Q, Q]\in\cM.
\e{322}
From \r{321} we have ($\d_a\equiv\d/\d\phi^a$)
\be
&&Q(0)=Q,\quad Q(\phi)\stackrel{\lea}{\d}_a i\hbar=[Q(\phi), f_a], \nn\\&&
-Q(\phi)\stackrel{\lea}{\d}_a
\stackrel{\lea}{\d}_b(i\hbar)^{2}(-1)^{\ve_a}=-[[Q(\phi),
f_a], f_b](-1)^{\ve_a}\equiv(f_a, f_b)_{Q(\phi)}
\e{323}
and the higher quantum antibrackets
\be
&&(f_{a_1},
f_{a_2},\ldots,f_{a_n})_{Q(\phi)}\equiv-Q(\phi)
\stackrel{\lea}{\d}_{a_1}\stackrel{\lea}{\d}_{a_2}\cdots
\stackrel{\lea}{\d}_{a_n}(i\hbar)^{n}(-1)^{E_n}=\nn\\&&=-[\cdots[[Q(\phi),
f_{a_1}],f_{a_2}],\cdots, f_{a_n}](-1)^{E_n},\quad E_n\equiv
\sum_{k=0}^{\left[{n-1\over 2}\right]}\ve_{a_{2k+1}}.
\e{28}
These are  quantum counterparts to the classical higher order antibrackets in
\cite{hanti}. (Condition
\r{251} may be written as
$(f_1, f_2, f_3)_Q=0$ for all $f_1, f_2, f_3\in \cM$. ) The correspondence
between the above quantum antibrackets and the corresponding classical
brackets is
symbolically
\be
&&(i\hbar)^{-n}(f_1,f_2,\ldots,f_n)_Q\; \lra\; (f_1,f_2,\ldots,f_n)
\e{29}
due to the commutator correspondence \r{30}. The Jacobi identities may be
derived
by differentiating $[Q(\phi), Q(\phi)]$ with respect to $\phi^a$ using \r{322}.
For instance,
\be
&&[Q(\phi), Q(\phi)]\stackrel{\lea}{\d}_{a}\stackrel{\lea}{\d}_{b}
\stackrel{\lea}{\d}_{c}(i\hbar)^{3}(-1)^{\ve_b+(\ve_a+1)(\ve_c+1)}\equiv 0
\e{324}
implies
\be
&&(f_a, (f_b, f_c)_{Q(\phi)})_{Q(\phi)}(-1)^{(\ve_a+1)(\ve_c+1)}+(f_c, (f_a,
f_b)_{Q(\phi)})_{Q(\phi)}(-1)^{(\ve_c+1)(\ve_b+1)}+\nn\\&&+(f_b, (f_c,
f_a)_{Q(\phi)})_{Q(\phi)}(-1)^{(\ve_b+1)(\ve_a+1)}=[[[[Q(\phi), f_a], f_b],
f_c], Q(\phi)](-1)^{\ve_a\ve_c}\equiv\nn\\&&\equiv[(f_a, f_b,
f_c)_{Q(\phi)}(-1)^{(\ve_a+1)(\ve_c+1)}, Q(\phi)].
\e{325}
which agrees with \r{271}.

For a nilpotent and hermitian $Q$ we may consistently define the quantum master
equation by
\be
&&Q|\phi\hb=0, \quad Q^2=0,
\e{33}
which in the wavefunction representation becomes a differential equation.
One may note that the quantum antibracket \r{24} between any two solutions
of \r{33} satisfies the property
\be
&&\vb\phi'|(f,
g)_Q|\phi\hb=\vb\phi'|\left(fQg-gQf(-1)^{(\ve_f+1)(\ve_g+1)}\right)|\phi\hb.
\e{34}
Thus, between $|\phi\hb$-states our first proposal \r{14} coincides with \r{24}.
Moreover, the Jacobi identities \r{11} for the quantum antibracket \r{24} are
satisfied for arbitrary operators between two $|\phi\hb$-states
\be
&&\vb\phi'|\left\{(f,(g,
h)_Q)_Q(-1)^{(\ve_f+1)(\ve_h+1)}+cycle(f,g,h)\right\}|\phi\hb=0, \quad \forall
f,g,h.
\e{340}
We may introduce a gauge-fixing fermion operator $\Psi$ in order to
define a physical matrix element of the form
\be
&&Z\equiv\vb \phi'|\exp{\{\hbar^{-2}[Q, \Psi]\}}|\phi\hb
\e{341}
between two $|\phi\hb$-states satisfying \r{33}. This matrix element is then
independent of $\Psi$. In fact, we have
\be
&&\del Z=\vb \phi'|\int_0^1d\al\,
\exp{\{\hbar^{-2}[Q, \Psi](1-\al)\}}\hbar^{-2}[Q,
\del\Psi]\exp{\{\hbar^{-2}[Q,
\Psi]\al\}}|\phi\hb=0
\e{342}
due to the master equation \r{33} and the hermiticity of $Q$.

We give now an explicit representation of $Q$.
Consider an antisymplectic manifold with Darboux coordinates $x^a$ and $x^*_a$,
$a=1,\ldots,n$, where $\ve(x^*_a)=\ve_a+1$, $\ve_a\equiv\ve(x^a)$. Consider
then these Darboux coordinates to be coordinates on a {\em symplectic}
manifold. The canonical coordinates of this symplectic manifold are then $\{x^a,
x^*_a, p_a, p^a_*\}$. We may now perform  a canonical quantization and choose
all operators which depends on $x^a$ and $x^*_a$ as the
class of commuting operators $\cM$. Finally, we  define the quantum
antibracket by
\r{26} with
$Q$ given by
\be
&&Q=p_a p^a_*(-1)^{\ve_a},
\e{36}
which is a nilpotent operator ($Q^2=0$). In this case we have
\be
&&(f, g)_Q=[f, [Q, g]]=[[f, Q], g]=\nn\\
&&=(-1)^{\ve_a}[f, p_a][p_*^a,
g]-(-1)^{\ve_a}[g, p_a][p^a_*,
f](-1)^{(\ve_f+1)(\ve_g+1)}.
\e{37}
The canonical commutation relations (the nonzero part),
\be
&&[x^a, p_b]=i\hbar\del^a_b,\quad [x^*_a, p^b_*]=i\hbar\del^b_a;\quad
p_a=-i\hbar\d_a(-1)^{\ve_a}, \quad p^a_*=i\hbar\d^a_*(-1)^{\ve_a},
\e{38}
imply then
\be
&&(i\hbar)^{-2}(f, g)_Q=f\stackrel{\lea}{\d}_a
\d_*^ag-g\stackrel{\lea}{\d}_a\d_*^a f(-1)^{(\ve_f+1)(\ve_g+1)}
\e{39}
in accordance with the correspondence \r{31}. The quantum master equation \r{33}
yields
\be
&&0=\vb x, x^*|Q|\phi\hb=-(i\hbar)^2\Delta\phi(x, x^*),
\e{40}
where $\Delta$ is the well-known nilpotent operator
\be
&&\Delta=(-1)^{\ve_a}{\d\over\d x^a}{\d\over\d x^*_a}.
\e{41}

In general coordinates $X^A=(x^a;x_a^*)$, $\ve(X^A)\equiv\ve_A$, the operator
\r{36} takes the form
\be
&&Q=-\half\rho^{-1/2}P_A\rho E^{AB}P_B\rho^{-1/2}(-1)^{\ve_B}, \quad Q^2=0,
\e{411}
where $E^{AB}(X)=-E^{BA}(X)(-1)^{(\ve_A+1)(\ve_B+1)}$ and $\rho(X)$ are the
antisymplectic metric and the volume form density respectively. $P_A$
satisfy
\be
&&[X^A, P_B]=i\hbar\del_B^A; \quad
P_A=-i\hbar\rho^{-1/2}\d_A\circ\rho^{1/2}(-1)^{\ve_A}, \quad \d_A\equiv
\d/\d X^A.
\e{412}
By means of \r{26} the quantum antibracket \r{39} generalizes here to
\be
&&(i\hbar)^{-2}(f, g)_Q=f\stackrel{\lea}{\d}_A E^{AB}\d_{B}\, g.
\e{413}
The nilpotency of $Q$ requires the tensor $E^{AB}$ to satisfy the cyclic
relation
\cite{BTU}
\be
&&E^{AD}\d_DE^{BC}(-1)^{(\ve_A+1)(\ve_C+1)}+cycle(A,B,C)=0,
\e{414}
which makes the antibracket \r{413} to satisfy the Jacobi identities, \ie
\r{325}
with vanishing right-hand side. If momenta $P_A$ are allowed to enter the
operator
$Q$ more than quadratically then a nonzero contribution appears on the
right-hand
side of \r{414}, and thereby  the right-hand side of \r{325} becomes nonzero as
well.

Finally we would like to mention that also the $Sp(2)$ version of the
BV-formalism
(see \cite{sp2}) may be ``quantized" in the above sense. The quantum $Sp(2)$
antibracket is defined by ($a=1,2$ is an $Sp(2)$-index.)
\be
&&(f, g)^a_Q=\half \left([f, [Q^a, g]]-[g, [Q^a,
f]](-1)^{(\ve_f+1)(\ve_g+1)}\right),
\e{42}
where $Q^a$ satisfies
\be
&&Q^{\{a}Q^{b\}}\equiv Q^{a}Q^{b}+Q^{b}Q^{a}=0.
\e{43}
Eq.\r{42} satisfies the relations
\be
&&[Q^{\{a}, (f, g)^{b\}}_Q]=([Q^{\{a}, f], g)^{b\}}_Q+(f, [Q^{\{a},
g])^{b\}}_Q(-1)^{\ve_f+1}=[[Q^{\{a},f], [Q^{b\}}, g]],
\e{431}
and for commuting operators \r{42} reduces to
\be
&&(f, g)^a_Q\equiv[f, [Q^a, g]]=[[f, Q^a], g], \quad \forall f, g\in\cM.
\e{432}
The Jacobi identities require
\be
&&[f, [g, [h, Q^a]]]=0, \quad \forall f,g,h\in\cM \quad \Leftrightarrow
\quad (f,
g)^a_Q\in\cM, \quad \forall f,g\in\cM.
\e{44}
The quantum master equations are here
\be
&&Q^a|\phi\hb=0, \quad a=1,2,
\e{46}
which are consistent due to \r{43}.\\ \\

\noindent
{\bf Appendix A:} \ {\bf Application to path integrals.}

The partition function $Z$, \ie the path
integral of the gauge fixed action, is within
the present scheme given by
$Z=\vb \Psi|S\hb$, where $|S\hb$ is the master state and $|\Psi\hb$ a gauge
fixing
state both satisfying the quantum master equation \r{33}, \ie
$Q|S\hb=Q|\Psi\hb=0$.
In the standard case where the hermitian $Q$ is given by \r{36} we have
explicitly
(cf. the Hamiltonian treatment in \eg \cite{BM})
\be
&&|S\hb=\exp{\left\{{i\hbar^{-1}}S(x,x^*)\right\}}|0\hb_{pp_*}, \quad
|\Psi\hb=\exp\left\{\hbar^{-2}[Q,
\Psi(x)]\right\}|0\hb_{px^*},
\e{b1}
where the operators have the hermiticity properties
\be
&&(x^a)^{\dag}=x^a,\;\; (x^*_a)^{\dag}=-x^*_a, \;\;
(p_a)^{\dag}=p_a(-1)^{\ve_a}, \;\;
(p^a_*)^{\dag}=p_*^a(-1)^{\ve_a},  \;\; \Psi^{\dag}=\Psi,
\e{b111}
 and where the vacuum states satisfy
\be
&&p_a|0\hb_{pp_*}=p_*^a|0\hb_{pp_*}=0, \quad
p_a|0\hb_{px^*}=x^*_a|0\hb_{px^*}=0, \quad Q|0\hb_{pp_*}=Q|0\hb_{px^*}=0.
\e{b2}
Note that $S(x,x^*)$ and $\Psi(x)$ belong to the class of commuting
operators. Note
also that
\be
&&Q|S\hb=0 \quad \Lra \quad \Delta \exp\left\{i\hbar^{-1}S(x,x^*)\right\}=0,
\e{b3}
and that $|\Psi\hb$ satisfies
\be
&& \left(x^*_a-\d_a\Psi(x)\right)|\Psi\hb=0, \quad
\left(p_a+p_*^b\d_b\d_a\Psi(x)\right)|\Psi\hb=0,
\e{b4}
which fixes $p_a$ and $x^*_a$. The gauge fixed partition function
becomes now
\be
&&Z=\vb \Psi|S\hb=\,_{px^*}\vb
0|\exp{\{\hbar^{-2}[ Q,
\Psi(x)]\}}\exp{\{i\hbar^{-1}S(x,x^*)\}}|0\hb_{pp_*}=\nn\\
&&=\int Dx Dx^*
\exp{\{i\hbar^{-1}S(x,x^*)\}}\del(x^*_a-\d_a\Psi(x)),
\e{b5}
where the last equality is obtained by inserting the completeness
relations\footnote{Note that for odd $n$ the states in \r{b6}, \r{b7} do
not have a
definite Grassmann parity \cite{RM}.}
\be
&&\int |x, x^*\hb D x D x^* \vb x, x^*|=\int  |x, p_*\hb D x D p_*\vb x,
p_*|=\bett,
\e{b6}
and the properties
\be
&&_{px^*}\vb 0|x, p_*\hb=\vb x,
x^*|0\hb_{pp_*}=1, \quad \vb
p_*|x^*\hb=(2\pi\hbar)^{-n_-/2}\hbar^{n_+/2}\exp{\{-i\hbar^{-1}p^a_*x^*_a\}},
\e{b7}
where $n_+$ ($n_-$) is the number of bosons (fermions) among the $x^a$
operators.
Eq.\r{b5} agrees with the standard BV quantization \cite{BV}. The
independence of
the gauge fixing operator $\Psi$ follows from \r{342}.
\\
\\

\noindent {\bf Appendix B:} \ {\bf Generalization to noncommuting
operators.}

 Consider the
operators $f_a$, $a=1,2,\ldots$, satisfying the nonabelian Lie algebra relations
\be
&&[f_a, f_b]=i\hbar U^c_{ab}f_c, \quad U^d_{ab}U^f_{dc}(-1)^{\ve_a\ve_c}+{
cycle}(a,b,c)\equiv 0.
\e{a1}
Define then a generating operator $Q(\phi)$ from a given $Q$ by \r{321}. We have
then
\be
&&Q(\phi)\stackrel{\lea}{\d}_a i\hbar=[Q(\phi),
\la^b_a(\phi)f_b(-1)^{\ve_a+\ve_b}],
\e{a2}
where integrability requires $\la^b_a$ to satisfy the Maurer-Cartan equation
\be
&&\d_a\la_b^c-\d_b\la_a^c(-1)^{\ve_a\ve_b}=\la^e_a\la^d_b
U^c_{de}(-1)^{\ve_b\ve_e+\ve_c+\ve_d+\ve_e},\quad \la^b_a(0)=\del^b_a.
\e{a3}
The explicit solution is
\be
&&\tilde{\la}=(1-e^{-X})/X, \quad
\tilde{\la}_b^a\equiv\la_b^a(-1)^{\ve_a(\ve_b+1)},
\;\; X_b^a\equiv\phi^cU^a_{cb}(-1)^{\ve_c+\ve_b(\ve_a+1)},
\e{a4}
where
$\phi^b\la_b^a=\phi^a$. In analogy with
\r{323} we define the generalized quantum antibracket by
\be
&&(f_a,
f_b)'_{Q(\phi)}\equiv-Q(\phi)\stackrel{\lea}{\d}_a
\stackrel{\lea}{\d}_b(i\hbar)^2(-1)^{\ve_a}=\nn\\
&&=-\la^c_a\la^d_b[[Q(\phi), f_d],
f_c](-1)^{\ve_b\ve_c+\ve_a}-i\hbar(\d_b\la^c_a)[Q(\phi),
f_c](-1)^{\ve_a(\ve_b+1)},
\e{a5}
and the corresponding higher antibrackets according to \r{28}.
Remarkably enough
 this
expression reduces exactly to the general ansatz \r{24} for $\phi^a=0$. However,
compared to the main text we have now a complete control over the Jacobi
identities
since
\r{322}   is valid leading to identities like \r{324}.  It should be noticed
that the class of operators satisfying
\r{a1} also includes all products of such operators: Define
$F_A$  to be all monomials of $f_a$,
\ie
$F_A\equiv f_a, f_af_b, f_af_bf_c,\ldots$. These operators do also satisfy a
nonabelian Lie algebra
$[F_A, F_B]=i\hbar\cU^C_{AB}F_C$.
In accordance with \r{321} we may therefore define
\be
&&Q(\Phi)\equiv
\exp{\{-(i\hbar)^{-1}F_A\Phi^A\}}\,Q\,\exp{\{(i\hbar)^{-1}F_A\Phi^A\}},
\e{a6}
where the parameters $\Phi^A$ are $\phi^a, \phi^{ab}, \phi^{abc}, \ldots$. Since
$F_A$ satisfies a nonabelian Lie
algebra we have also here integrable equations of the
form
\r{a2} and generalized quantum antibrackets (${\d}_A=\d/\d\Phi^A$)
\be
&&(F_A, F_B)'_{Q(\Phi)}\equiv
-Q(\Phi)\stackrel{\lea}{\d}_A\stackrel{\lea}{\d}_B(i\hbar)^2(-1)^{\ve_A}, \quad
\ve_A\equiv
\ve(F_A)=\ve(\Phi^A),
\e{a7}
which again reduces to \r{24} at $\Phi^A=0$. Thus, this definition of
antibrackets
for  $f_a$-operators and their arbitrary products
 agrees exactly with what we should have.  Furthermore,  we have complete
control
over all the Jacobi identities, since
\r{322}   is valid even for \r{a6}, as well as the consistent Leibniz' rules.
 The Leibniz' rule 5) requires \eg the following additional conditions from
\r{25}
\be
&&(f_af_b, f_c)_Q-f_a(f_b, f_c)_Q-(f_a,
f_c)_Qf_b(-1)^{\ve_b(\ve_c+1)}=\nn\\ &&=\half\left([f_a, Q]i
\hbar U_{bc}^d f_d(-1)^{\ve_b} + f_d
U_{ac}^d i\hbar[Q, f_b](-1)^{(\ve_b+\ve_d+1)(\ve_c+1)+\ve_a\ve_d}\right)
=\nn\\&&=i\hbar\half\left([f_a, Q] U_{bc}^d-[f_b, Q]
U_{ac}^d(-1)^{(\ve_a+1)(\ve_b+1)}\right)f_d(-1)^{\ve_b}+
\nn\\&&+i\hbar\half U_{ac}^d[f_d,
[f_b, Q]](-1)^{\ve_c(\ve_b+1)}=0.
\e{a8}

The above construction may be naturally generalized to operators $f_a$
satisfying the
nonabelian algebra in \r{a1} with operator-valued structure coefficients
$U^c_{ab}$,
which  commute among themselves, but not with $f_a$ in such a way that $[[f_d,
U^c_{ab}], U^g_{ef}]=0$. The Jacobi identities in
\r{a1} are then modified as
\be
&& \left(i\hbar U^d_{ab}U^e_{dc}+[U^e_{ab},
f_c](-1)^{\ve_c\ve_e}\right)(-1)^{\ve_a\ve_c}+{ cycle}(a,b,c)\equiv 0.
\e{a81} In this case
there exist  solutions  $\la^b_a(\phi)$  commuting among themselves and with
$U^c_{ab}$. In fact, the Maurer-Cartan equation \r{a3} is  valid for the
commuting
operators
$V\la_a^bV^{-1}$ and $VU_{ab}^cV^{-1}$ where
$V(\phi)\equiv\exp{\{(i\hbar)^{-1}f_a\phi^a\}}$.\\
\\

\noindent {\bf Appendix C:} \ {\bf A group theoretic application.}

An interesting group theoretic application is obtained if we consider
$\la^b_a(\phi)$  satisfying the Maurer-Cartan equation \r{a3} with the parameter
$\phi^a$ replaced by  the commuting coordinate operator $x^a$.
$\mu^b_a(x)\equiv(\la^{-1})^b_a(x)$ satisfies then
\be
&&\mu_a^c\d_c\mu_b^d-\mu_b^c\d_c\mu^d_a(-1)^{\ve_a\ve_b}
=U^c_{ab}(-1)^{\ve_a+\ve_b+\ve_c}\mu^d_c.
\e{a9}
In terms of $\mu^b_a(x)$, $x^*_a$, $p_a$, and $p_*^a$, we have then the
nilpotent
$Q$-operator (cf. \cite{AD})
\be
&Q=&\left(\mu^b_a(x)\left(p_b-{i\hbar\over
2}\stackrel{\lea}{\d}_b\right)+{i\hbar\over
2}U^b_{ab}(-1)^{\ve_a}\right)(-1)^{\ve_b}p_*^a+\nn\\
&&+\half
x^*_cU^c_{ab}p^b_*p^a_*(-1)^{(\ve_b+1)(\ve_c+1)+\ve_a\ve_c},
\e{a10}
where the $\hbar$-terms are required by hermiticity. This is a particular
example of the general expression \r{411}. Since
\r{a10} is at most quadratic in the momenta, we may consistently define the
quantum
antibracket by the formula \r{26} for commuting operators. Calculating the
double
commutator we find for  operators $f, g$, which are functions of $x^a$ and
$x^*_a$
\be
&&(i\hbar)^{-2}(f, g)_Q=-f\stackrel{\lea}{\d^a_*}\left(\mu^b_a\d_b+U_{ab}^c
x^*_c(-1)^{\ve_a+\ve_b+\ve_c}\d_*^b\right)
 g- (f\lra g)(-1)^{(\ve_f+1)(\ve_g+1)}.\nn\\
\e{a11}
In particular we have
\be
&&(i\hbar)^{-2}(x^a, x^b)_Q=0,\quad (i\hbar)^{-2}(x^*_a,
x^b)_Q=-\mu^b_a(x), \nn\\
&&(i\hbar)^{-2}(x^*_a, x^*_b)_Q=-U^c_{ab}(-1)^{\ve_a+\ve_b+\ve_c}x^*_c.
\e{a12}
The
Jacobi identities are satisfied due to \r{a9}.

From \r{a10} we may define constraint operators $T_a$ by
\be
&T_a\equiv& -(i\hbar)^{-1}[Q, x^*_a]=-\mu^b_a\left(p_b-{i\hbar\over
2}\stackrel{\lea}{\d}_b\right)(-1)^{\ve_a+\ve_b}-\nn\\&&-{i\hbar\over
2}U_{ab}^b(-1)^{\ve_b}-x^*_cU^c_{ab}p^b_*(-1)^{(\ve_a+\ve_b+1)(\ve_c+1)}.
\e{a13}
$T_a$ and $T^*_a\equiv - x^*_a(-1)^{\ve_a}$ satisfy the commutator algebra
\be
&&(i\hbar)^{-1}[T_a, T_b]=U_{ab}^c T_c, \quad (i\hbar)^{-1}[T_a, T^*_b]=U_{ab}^c
T^*_c,\nn\\
&&(i\hbar)^{-1}[T^*_a, T^*_b]=0; \quad  (i\hbar)^{-1}[T^*_a, Q]=T_a, \quad
(i\hbar)^{-1}[T_a, Q]=0,
\e{a14}
and in terms of the antibracket \r{24} with the $Q$-operator \r{a10} we
have their
dual algebra
\be
&&(i\hbar)^{-2}(T^*_a, T^*_b)_Q=U_{ab}^c T^*_c, \quad (i\hbar)^{-2}(T^*_a,
T_b)_Q=\half U_{ab}^c T_c,\nn\\
&&(i\hbar)^{-2}(T_a, T_b)_Q=0;  \quad
(i\hbar)^{-2}(T_a, Q)_Q=0, \quad (i\hbar)^{-2}(T^*_a, Q)_Q=0.
\e{a15}\\ \\

\noindent
{\bf Acknowledgments}

I.A.B. would like to thank Lars Brink for his very warm hospitality at the
Institute of Theoretical Physics, Chalmers and G\"oteborg University. The
authors would like to thank A. M. Semikhatov for sending us, prior to
publication, a very interesting paper \cite{AMS} devoted to a related
problem. The
work is partially supported by INTAS-RFBR grant 95-0829. The work of I.A.B.
is also
supported by INTAS grant 96-0308 and by RFBR grants 96-01-00482, 96-02-17314.
I.A.B. and R.M. are thankful to the Royal Swedish Academy of Sciences for
financial support.


\begin{thebibliography}{Simple}



\bibitem{anti}
J. A. Schouten, \ {\sl Koninklijke Nederlandse Akademie van Wetenschappen,
Series
A, Proceedings}\ {\bf 2}, 449 (1940); {\em On the differential operators of
first
order in tensor calculus}, in Convegno Internazionale di Geometria
Differenziale,
Italy, Sept. 20-26, 1923, Edizioni Cremonese delle Casa Editrice Perrella, Rome,
(1954);\\
C. Buttin, \ {\sl C.r. Acad. Sci. Paris}\ {\bf 269}, A87 (1969);\\
 J. Zinn-Justin, {\em Renormalization of
gauge theories}, in Trends in Elementary
Particle Theory, Lecture Notes in Physics (eds. Rollnik
and Dietz), vol 37, Springer-Verlag,
Berlin, 1975;\\
E. Witten,  \ {\sl Modern Phys. Lett.}\ {\bf A5}, 487
(1990).

\bibitem{BV}I. A. Batalin, and G. A. Vilkovisky, \ {\sl Phys. Lett.}\ {\bf
102B}, 27
(1981); \ {\sl Phys. Rev.}\ {\bf D28}, 2567
(1983)

\bibitem{Panti}D. V. Volkov, A. I. Pashnev,  V. A. Soroka, and V. I. Tkach,
\ {\sl
JETP Lett.} \ {\bf 44}, 70 (1986); \ {\sl Teor. Mat. Fiz.} \ {\bf 79}, 117
(1989);\\ V. A. Soroka, \ {\sl Lett. Math. Phys.}\ {\bf 17}, 201 (1989).


\bibitem{KN}O. M. Khudaverdian,\ {\sl J. Math. Phys. }\ {\bf 32}, \ 1934
(1991);\\
O. M. Khudaverdian and A. P. Nersessian,\ {\sl J. Math. Phys. }\ {\bf 32},
\ 1938
(1991).

\bibitem{quanti}D. V. Volkov, V. A. Soroka, and V. I. Tkach, \ {\sl Yad. Fiz.}\
{\bf 44}, 810 (1986);\\
D. V. Volkov, and V. A. Soroka, \ {\sl Yad. Fiz.}\
{\bf 46}, 110 (1987);\\
V. A. Soroka, \ {\sl JETP Lett.}\
{\bf 59}, 219 (1994).

\bibitem{AMS}M. A. Grigoriev, A. M. Semikhatov, and I. Yu. Tipunin,
 \ {\em Gauge Symmetries of the Master Action}, \ {\tt hep-th/9804156}

\bibitem{hanti}J.-L. Koszul, Ast\'erisque, hors serie 257 (1985);\\
F. Akman, {\tt q-alg/9506027}\\
K. Bering, P. H. Damgaard,  and  J. Alfaro \ {\sl Nucl. Phys. }\
{\bf B478}, 459 (1996)\\
I. A. Batalin, K. Bering, and P. H. Damgaard, \ {\sl Phys. Lett.}\ {\bf
389B}, 673
(1996)


\bibitem{BTU} I.A.~Batalin,
and I.V.~Tyutin,
 \ {\sl Int. J. Mod. Phys.} \ {\bf A8}, 2333 (1993)

\bibitem{sp2} I.A.~Batalin,
P.M.~Lavrov,  and I.V.~Tyutin,
\JMP\ 31 (1990) 1487;\\
{\em ibid}\ 32 (1990) 532;
{\em ibid}\ 32 (1990) 2513;\\
I. A. Batalin,
R. Marnelius and A. M. Semikhatov,
 \ {\sl Nucl. Phys.}\ {\bf
            B446}, \ 249 (1995);
I. A. Batalin and
R. Marnelius,
 \ {\sl Nucl. Phys.}\ {\bf
            B465}, \ 521 (1996)


\bibitem{BM} I. A. Batalin and
R. Marnelius,
 \ {\sl Nucl. Phys.}\ {\bf
            B442}, \ 669 (1995)

\bibitem{RM}
R. Marnelius,
 \ {\sl Int. J. Mod. Phys.}\ {\bf
            A5}, \ 329 (1990)

\bibitem{AD}J. Alfaro and P. H. Damgaard, \ {\sl Phys. Lett.}\ {\bf 369B}, 289
(1996)


\end{thebibliography}
\end{document}